\documentclass[prb, superscriptaddress,twocolumn,10pt,
 amsmath,
 amssymb,
 aps,
 longbibliography
]{revtex4-2}

\usepackage{graphicx}
\usepackage[colorlinks,linkcolor=blue,citecolor=blue,urlcolor=blue]{hyperref}
\usepackage{stix}
\usepackage[utf8]{inputenc} 
\usepackage[T1]{fontenc}
\usepackage{booktabs}

\begin{document}

\renewcommand{\vec}[1]{\mathbf{#1}}
\newcommand{\gvec}[1]{\boldsymbol{#1}}
\newcommand{\en}{\varepsilon}

\newcommand{\red}[1]{\textcolor{red}{#1}}
\newcommand{\orange}[1]{\textcolor{orange}{#1}}
\newcommand{\Markus}[1]{\textcolor{blue}{Markus: #1}}

\newcommand{\up}[1]{\textcolor{OliveGreen}{#1}}
\newcommand{\ms}[1]{\textcolor{WildStrawberry}{[#1]}}

\title{First-principle tight-binding approach to angle-resolved photoemission spectroscopy simulations: importance of light-matter gauge and ubiquitous interference effects }%

\author{Yun Yen}
\affiliation{Laboratory for Materials Simulations, Paul Scherrer Institute, CH-5232 Villigen PSI, Switzerland}
\affiliation{École Polytechnique Fédérale de Lausanne (EPFL),  CH-1015 Lausanne, Switzerland}
\author{Gian Parusa}
\affiliation{Laboratory for Materials Simulations, Paul Scherrer Institute, CH-5232 Villigen PSI, Switzerland}
\affiliation{Department of Physics, University of Fribourg, CH-1700 Fribourg, Switzerland}
\author{Michael Sch\"uler}
\email{michael.schueler@psi.ch}
\affiliation{Laboratory for Materials Simulations, Paul Scherrer Institute, CH-5232 Villigen PSI, Switzerland}
\affiliation{Department of Physics, University of Fribourg, CH-1700 Fribourg, Switzerland}

\date{\today}

\begin{abstract}
    Angle-resolved photoemission spectroscopy (ARPES) is one of the most powerful techniques to study the electronic structure of materials. To go beyond the paradigm of band mapping and extract aspects of the Bloch wave-functions, the intricate interplay of experimental geometry, crystal structure, and photon polarization needs to be understood. In this work we discuss several model approaches to computing ARPES signals in a unified fashion. 
    While we represent the Bloch wave-functions by first-principle Wannier functions, we introduce different approximations to the final states and discuss the implications for the predictive power. We also introduce various light-matter gauges and explain the role of the inevitable breaking of gauge invariance.
    Finally, we benchmark the different models for the two-dimensional semiconductor WSe$_2$, known for its strong Berry curvature, orbital angular momentum (OAM), and nontrivial orbital texture. The models are compared based on their ability to simulate photoemission intensity and interpret circular dichroism in ARPES (CD-ARPES). 
    We show that interference effects are crucial to understanding the circular dichroism, and explain their photon-energy dependence. Our in-depth analysis provides insights into the advantages and limitations of various model approaches in clarifying the complex interplay between experimental observables and underlying orbital texture in materials.
\end{abstract}

\maketitle

\section{Introduction}

The rise of materials with nontrival topological properties~\cite{hasan_colloquium:_2010} was paralleled and driven by the advances of angle-resolved photoemission spectroscopy (ARPES)~\cite{schonhense_multidimensional_2018,sobota_angle-resolved_2021,rader_angle-resolved_2021,lv_angle-resolved_2019,dil_spin-_2019,lu_angle-resolved_2012}. This experimental techniques provides direct access to the band structure of the occupied Bloch states in periodic solids. More recently, pump-probe time-resolved ARPES (trARPES) extended the capability to map out also the excited states~\cite{boschini_role_2020}. 

Beyond extracting the band energies $\varepsilon_\alpha(\vec{k})$, the focus has shifted more recently towards accessing aspects of the underlying wave-function $\psi_{\vec{k}\alpha}(\vec{r})$. In particular, the intricate dependence of the spectra on the polarization of the photons and the experimental geometry has been shown to contain fingerprints of the wave-function~\cite{hwang_direct_2011,beaulieu_unveiling_2021,beaulieu_revealing_2020-1,unzelmann_momentum-space_2021,yen_controllable_2023}. The most prominent example is circular dichroism in ARPES (CD-ARPES)~\cite{schonhense_circular_1990,fedchenko_4d_2019}. For simple systems, a link between the circular dichroism in the angular distribution (CDAD) and the orbital angular momentum (OAM) can be established~\cite{park_detecting_2012,gierz_graphene_2012,wang_circular_2013,cho_experimental_2018,cho_studying_2021,schuler_local_2020,schuler_how_2020}. For more complex systems however, the extraction of the precise orbital texture is difficult due to a number of complicating effects. 
Multiple interference of the partial waves originating from the same and from different atoms give rise to pronounced dependence on the photon energy~\cite{kern_simple_2023,gierz_illuminating_2011,heider_geometry-induced_2023} while the scattering of the photoelectrons off the lattice imprint a complex energy dependence onto the final states of the photoemission process~\cite{strocov_high-energy_2023}. 

While first-principle methods~\cite{krasovskii_calculation_1999,krasovskii_band_2007,ebert_calculating_2011,de_giovannini_first-principles_2017,ono_surface_2021} -- akin numerical experiments -- are available and can help understand experiments, a more transparent method would be beneficial to clarify the connection between spectra and the Bloch wave-functions. To this end, various model approaches have been employed with the goal of disentangling the contributions to the photoemission signal~\cite{moser_experimentalists_2017,day_computational_2019,schuler_polarization-modulated_2022,feidt_development_2019,schuler_probing_2022,moser_toy_2023}. However, potentially uncontrolled approximations, inequivalent schemes of incorporating the light-matter coupling, and unnecessary assumptions make it unclear which model provides the best compromise of accuracy and transparency. 

In this paper we tackle this problem by a systematic assessment of the different approximation schemes. We start by a rigorous derivation, where we introduce the different light-matter gauges. In particular, we employ the modern theory of polarization to construct the dipole operator~\cite{resta_quantum-mechanical_1998} from a quantum geometrical perspective.
The Bloch wave-functions $\psi_{\vec{k}\alpha}(\vec{r})$ of the occupied bands are obtained from first-principle density-function theory (DFT), whereas the photoelectron states are treated approximately.

Finally, we benchmark the different models against each other for the paradigmatic two-dimensional semiconductor WSe$_2$, which exhibits strong Berry curvature, OAM, and nontrivial orbital texture in the valence bands. 
This system allows us to display the advantage of the model approach to ARPES, as we can analyze the various contributions to the simulated photoemission intensity. In particular, we discuss the importance of interference effects in the photoemission process, which gives rise to a pronounced photon-energy dependence of the CDAD and a complex texture. We then discuss to what extend the CDAD can be linked to the OAM under typical experimental conditions.

\section{Wannier function approach}
\label{sec:wannier_method}

We consider a 2D material or the surface of a bulk material, described by the effective single-particle Hamiltonian
\begin{align}
    \label{eq:ham1} 
    \hat{H} = -\frac{\nabla^2}{2} + \hat{V} \ .
\end{align}
Here, and it what follows, we use atomic units (a.u.) consistently unless stated otherwise.
Typically, $\hat{H}$ is taken as the Kohn-Sham Hamiltonian within the framework of density-functional theory (DFT), but other choices such as Hartree-Fock are possible. For the sake of clarity we assume the potential energy operator $\hat{V}$ represents a local potential $\hat{V} \rightarrow V(\vec{r})$ (the generalization to non-local potentials is straightforward). The potential is periodic in the plane parallel to the surface, $V(\vec{r} + \vec{R}) = V(\vec{r})$, with $\vec{R}$ denoting the in-plane lattice vectors. According to the Bloch theorem, the eigenstates of the Hamiltonian~\eqref{eq:ham1} are the Bloch states $\hat{H} |\psi_{\vec{k}\alpha}\rangle$ with the Bloch-phase condition $\psi_{\vec{k}\alpha}(\vec{r}+ \vec{R}) = e^{i \vec{k}\cdot \vec{R}} \psi_{\vec{k}\alpha}(\vec{r})$. The quasi-momentum $\vec{k}$ is a vector from the 2D Brillouin zone (BZ). While it seems that this formulation is geared towards 2D materials,  bulk materials are also captured by considering a slab geometry, where the information about the layers translates into, in principle infinitely, many bands $\alpha$. 

\subsection{Bloch states: bound states and photoelectron states}

The Bloch states $|\psi_{\vec{k}\alpha}\rangle$ can be grouped into bound states with a discrete band index $\alpha$, and photoelectron states (also known as inverse LEED states) in the continuum. For bound states, we can employ the Wannier representation 
\begin{align}
    \label{eq:wannier1}
    \psi_{\vec{k}\alpha}(\vec{r}) = \frac{1}{\sqrt{N}} \sum_{\vec{R}j} e^{i\vec{k}\cdot \vec{R}}C_{j\alpha}(\vec{k}) \phi_j(\vec{r} - \vec{r}_j - \vec{R}) \ .
\end{align}
Here, $\phi_j(\vec{r})$ denote the Wannier orbitals, which are localized at the Wannier center $\vec{r}_j$. Within empirical TB approaches such as the Slater-Koster theory, the Wannier orbitals can be approximated by atom-like orbitals centered at the atomic sites $\vec{r}_j$. The Wannier orbitals can also be constructed from the Bloch states $\psi_{\vec{k}\alpha}(\vec{r})$ by the Wannierization procedure, delivering projective or maximally localized Wannier functions~\cite{marzari_maximally_2012}. The other ingredients in Eq.~\eqref{eq:wannier1} are the number of lattice sites $N\rightarrow \infty$, and the rotation coefficients $C_{j\alpha}(\vec{k})$ which contain the amplitude of Wannier orbital $j$ in the band $\alpha$.

The Hamiltonian~\eqref{eq:ham1} also hosts photoelectron states, which we label as $|\chi_{\vec{p}}\rangle$ with the three-dimensional momentum $\vec{p} = \vec{p}_\parallel + p_\perp \hat{z}$ and energy $\en_{\vec{p}} = \vec{p}^2 / 2$. The photoelectron states also obey the Bloch phase condition $\chi_{\vec{p}}(\vec{r} + \vec{R}) = e^{i \vec{p}_\parallel \cdot \vec{R}} \chi_{\vec{p}}(\vec{r})$. Computing $|\chi_{\vec{p}}\rangle$ from first principles is challenging, but possible within the one-step approach to ARPES. 

\subsection{Fermi's Golden Rule \label{subsec:fermi}}

With the eigenstates of the Hamiltonian defined, we can compute the photoemission intensity from Fermi's Golden Rule 
\begin{align}
    \label{eq:fermi_golden}
    I(\vec{p}) = \sum_{\vec{k}\alpha} f_\alpha(\vec{k}) |\langle \chi_{\vec{p}}| \hat{\Delta} | \psi_{\vec{k}\alpha} \rangle | ^2 \delta(\varepsilon_\alpha(\vec{k}) + \omega - \en_{\vec{p}} ) \ .
\end{align}
Here, $f_{\alpha}(\vec{k})$ denotes the occupation of the Bloch state $| \psi_{\vec{k}\alpha} \rangle $ (band energy $\varepsilon_\alpha(\vec{k})$). The Dirac delta function in Eq.~\eqref{eq:fermi_golden}
ensures the energy conservation of the photoemission process. The light-matter coupling is incorporated through the operator $\hat{\Delta}$. Within the minimal coupling scheme, it is given by
\begin{align}
    \label{eq:velocity_gauge}
    \hat{\Delta} = \vec{e}\cdot\hat{\vec{p}} \ ,
\end{align}
where $\vec{e}$ denotes the polarization vector. 

The Bloch-phase condition for both initial and final states in the matrix element in Eq.~\eqref{eq:fermi_golden} implies the conservation of the in-plane momentum, as 
\begin{align*}
    \langle \chi_{\vec{p}}| \hat{\Delta} | \psi_{\vec{k}\alpha} \rangle &= \int d\vec{r}\, \chi^*_{\vec{p}}(\vec{r}) \hat{\Delta}  \psi_{\vec{k}\alpha}(\vec{r}) \\ &= \sum_{\vec{R}}   \int_\mathrm{UC} d\vec{r}\, \chi^*_{\vec{p}}(\vec{r}+\vec{R}) \hat{\Delta}  \psi_{\vec{k}\alpha}(\vec{r}+\vec{R}) \\ &=  \sum_{\vec{R}} e^{i(\vec{k} - \vec{p}_\parallel)\cdot \vec{R}} \int_\mathrm{UC} d\vec{r}\, \chi^*_{\vec{p}}(\vec{r}) \hat{\Delta}  \psi_{\vec{k}\alpha}(\vec{r}) \ .
\end{align*}
Here, $\int_\mathrm{UC}$ denotes the integration over the 2D unit cell. The sum over $\vec{R}$ in the last line yields $N \delta_{\vec{k} + \vec{G},\vec{p}_\parallel }$ with a reciprocal lattice vector $\vec{G}$. Focusing on the first BZ ($\vec{G}=0)$, we can thus re-label $|\chi_{\vec{p}}\rangle \rightarrow |\chi_{\vec{k},E}\rangle $ and simplify Fermi's Golden rule~\eqref{eq:fermi_golden} to
\begin{align}
    \label{eq:fermi_golden_2}
    I(\vec{k},E) = \sum_{\alpha} f_\alpha(\vec{k}) |M_{\alpha}(\vec{k},E)|^2 \delta(\varepsilon_\alpha(\vec{k}) + \omega -E ) \ .
\end{align}
Here, $E$ denotes the energy of the photoelectron $E = (\vec{k}^2 + p^2_\perp)/2$, while $M_{\alpha}(\vec{k},E) =\langle \chi_{\vec{k},E}| \hat{\Delta} | \psi_{\vec{k}\alpha} \rangle $ stands for the photoemission matrix element.

The advantage of the Wannier representation~\eqref{eq:wannier1} displays itself in the transparent exansion of the matrix elements. Exploiting the momentum conservation, one finds
\begin{align}
    \label{eq:mel_expansion_1}
    M_{\alpha}(\vec{k},E) = \sqrt{N}\sum_{j} C_{j\alpha}(\vec{k}) M^\mathrm{orb}_j(\vec{k},E) 
\end{align}
Eq.~\eqref{eq:mel_expansion_1} allows us to interpret the photoemission matrix element as a coherent sum of \emph{orbital} matrix elements 
\begin{align}
    \label{eq:mel_orbital_mom}
    M^\mathrm{orb}_j(\vec{k},E) =\int d\vec{r}\, \chi^*_{\vec{k},E}(\vec{r} + \vec{r}_j) \hat{\Delta} \phi_j(\vec{r}) 
\end{align} 
in terms of the Wannier orbitals. Whenever a band $\alpha$ is dominant by a particular orbital character, only one coefficient $C_{j\alpha}(\vec{k})$ enters the sum in Eq.~\eqref{eq:mel_expansion_1}. In this case, the symmetries of the relevant orbital directly manifest in the photoemission matrix elements. In contrast, if several coefficients $C_{j\alpha}(\vec{k})$ contribute, interferences of the different orbital channels complicate the photoemission signal.

While we have employed a particular choice for the light-matter coupling~\eqref{eq:velocity_gauge} directly following the minimal coupling scheme (known as velocity gauge), Fermi's Golden rule remains valid for other choices.
As long as the initial states $|\psi_{\vec{k}\alpha}\rangle$ and the final states $|\chi_{\vec{k},E}\rangle$ are eigenstates of the same Hamiltonian~\eqref{eq:ham1}, gauge-invariance is guaranteed. Choosing the light-matter coupling operator $\hat{\Delta} = \vec{e}\cdot \hat{\vec{r}}$
(dipole gauge) or $\hat{\Delta} = \vec{e}\cdot \nabla V(\vec{r})$
(acceleration gauge) yields (up to a prefactor) the identical expression for the intensity~\eqref{eq:fermi_golden_2}. Additional approximations break the gauge invariance, rendering the choice of the gauge an important step in practical calculations.

\subsection{Dipole gauge}

In the dipole gauge the light-matter coupling is represented by $\hat{\Delta} = \vec{e}\cdot\hat{\vec{r}}$. While the dipole operator $\hat{\vec{r}}$  itself is ill-defined for periodic solids, the matrix elements with respect to different Bloch states is well-defined~\cite{resta_quantum-mechanical_1998,resta_electrical_2010}:
\begin{align}
    \label{eq:berry1}
    \langle \psi_{\vec{k}\alpha} | \hat{\vec{r}} | \psi_{\vec{k}\alpha^\prime} \rangle = i \langle u_{\vec{k}\alpha} | \nabla_{\vec{k}} u_{\vec{k}\alpha^\prime} \rangle \quad \mathrm{for} \quad  \alpha \ne \alpha^\prime \ ,
\end{align}
where $u_{\vec{k}\alpha}(\vec{r}) = e^{-i \vec{k}\cdot\vec{r}} \psi_{\vec{k}\alpha}(\vec{r})$ denote the cell-periodic Bloch states.
Eq.~\eqref{eq:berry1} establishes a link between the inter-band dipole matrix element and quantum geometric properties: the right-hand side of Eq.~\eqref{eq:berry1} is nothing else than the Berry connection $\vec{A}_{\alpha \alpha^\prime}(\vec{k})$.

Since the photoelectron states $|\chi_{\vec{k},E}\rangle$ are eigenstates of $\hat{H}$, we can interpret the photoemission matrix element as an inter-band dipole matrix element. Separating the polarization in the in-plane ($\vec{e}_\parallel$) and out-of-plane ($e_\perp$) components, one finds
\begin{align}
    \label{eq:mel_dip_1}
    M_{\alpha}(\vec{k},E) &= i\vec{e}_\parallel \cdot \langle \tilde{\chi}_{\vec{k},E} | \nabla_{\vec{k}} u_{\vec{k}\alpha} \rangle + e_\perp \langle \chi_{\vec{k},E} | z | \psi_{\vec{k}\alpha} \rangle \ ,
\end{align}
where $\tilde{\chi}_{\vec{k},E}(\vec{r}) = e^{-i \vec{k}\cdot \vec{r}} \chi_{\vec{k},E}(\vec{r}) $ is the cell-periodic part of photoelectron state. Note that the broken periodicity in the $z$ direction renders the corresponding dipole operator the same as for finite systems. For evaluating the first term on the right-hand side of Eq.~\eqref{eq:mel_dip_1}, we employ the Wannier representation~\eqref{eq:wannier1} for the cell-periodic Bloch wave-function. It is convenient for this purpose to re-define the switch the convention of the coefficients, $C_{j\alpha}(\vec{k}) = e^{i\vec{k}\cdot \vec{r}_j} \tilde{C}_{j\alpha}(\vec{k})$. Taking the $\vec{k}$-derivative of $u_{\vec{k}\alpha}(\vec{r})$ and exploiting the translational invariance, one then finds the dipole matrix element
\begin{widetext}
    \begin{align}
    \label{eq:dipole_matrix}
    \vec{D}_\alpha(\vec{k},E) \equiv i\langle \tilde{\chi}_{\vec{k},E} | \nabla_{\vec{k}} u_{\vec{k}\alpha} \rangle &=
    \sqrt{N}\sum_j e^{i\vec{k}\cdot \vec{r}_j}\tilde{C}_{j\alpha}(\vec{k}) \int d\vec{r}\, \chi^*_{\vec{k},E}(\vec{r}) (\vec{r} - \vec{r}_j) \phi_j(\vec{r} - \vec{r}_j) \nonumber \\ & \quad +  i \sqrt{N}\sum_j e^{i\vec{k}\cdot \vec{r}_j} \nabla_{\vec{k}} \tilde{C}_{j\alpha}(\vec{k}) \int d\vec{r}\, \chi^*_{\vec{k},E}(\vec{r})\phi_j(\vec{r} - \vec{r}_j) \ .
\end{align}
\end{widetext}
The first contribution to the dipole matrix elements~\eqref{eq:dipole_matrix} is analogous to the coherent sum~\eqref{eq:mel_expansion_1} of orbital matrix elements~\eqref{eq:mel_orbital_mom}. We refer to it as atom-centered contribution $\vec{D}^\mathrm{AC}(\vec{k},E)$, whereas the second term on the right-hand side of Eq.~\eqref{eq:dipole_matrix} originates from the quantum geometry of the Bloch states as encoded in the $\vec{k}$-dependence of the coefficients $\tilde{C}_{j\alpha}(\vec{k})$. While such quantum geometric contributions to dipole matrix elements play an important role in linear and nonlinear excitation phenomena in solids~\cite{ahn_riemannian_2021,von_gersdorff_measurement_2021,baykusheva_strong-field_2021}, for in example in transport and high-harmonic generation, they typically play only a minor role in the photoemission process. First of all, the $\vec{k}$-derivative acts only on the in-plane coordinates, 
which means that quantum geometric contributions are only relevant for in-plane photon polarization. Furthermore, the magnitude of the second term in Eq.~\eqref{eq:dipole_matrix} is small compared to $\vec{D}^\mathrm{AC}(\vec{k},E)$ unless the coefficients $\tilde{C}_{j\alpha}(\vec{k})$ exhibit a strong momentum dependence.  

In the following, we will refer to $\vec{D}_\alpha(\vec{k},E) \approx \vec{D}^\mathrm{AC}(\vec{k},E)$ as the atom-centered approximation (ACA). As we will show below, the ACA as presented in this section provides the formal foundation for the ad-hoc approaches often encountered in the literature~\cite{park_orbital_2012,day_computational_2019,devereaux_web-based_2021}.

Exploiting the orthogonality of photoelectron and bound states, we can simplify the photoemission matrix element within the ACA as
\begin{align}
    \label{eq:mel_dip_aca}
    M^\mathrm{ACA}_\alpha(\vec{k}, E) = \sqrt{N} \sum_j C_{j\alpha}(\vec{k}) M^{\mathrm{orb,ACA}}_j(\vec{k},E)
\end{align}
with
\begin{align}
    \label{eq:mel_dip_aca_orb}
    M^{\mathrm{orb,ACA}}_j(\vec{k},E) = \int d\vec{r}\, \chi^*_{\vec{k},E}(\vec{r} + \vec{r}_j) 
    \vec{e}\cdot \vec{r}
    \phi_j(\vec{r})  \ . 
\end{align}
Eqs.~\eqref{eq:mel_dip_aca}--\eqref{eq:mel_dip_aca_orb} are identical to the matrix elements within the minimal coupling scheme~\eqref{eq:mel_expansion_1}--\eqref{eq:mel_orbital_mom} upon replacing $\hat{\Delta}\rightarrow \vec{e}\cdot \vec{r}$. However, we stress that gauge invariance is broken when employing the ACA, and velocity gauge and dipole gauge will yield different results.

\subsection{Plane-wave approximation}

Up to this point, we have assumed exact photoelectron states, i.\,e. $\hat{H}|\chi_{\vec{k},E}\rangle = E |\chi_{\vec{k},E}\rangle$. However, explicitly calculating $|\chi_{\vec{k},E}\rangle$ is challenging. Finding approximations to the photoelectron states that balance accuracy and simplicity is the practical path forward that we will take here. 

The simplest possible approximation is the plane-wave approximation (PWA): $\chi_{\vec{k},E}(\vec{r})\approx e^{i\vec{p}\cdot \vec{r}}$. The PWA is based on the fact that at high energy $E$ the crystal potential plays only a minor role, rendering $|\chi_{\vec{k},E} \rangle$ an approximate eigenstate of the kinetic energy only. While the PWA generally becomes more accurate at large $E$, this trend is not monotonic~\cite{strocov_high-energy_2023}. Assuming the PWA to be applicable for now, we can make use of the property
\begin{align}
    \label{eq:pw_shift}
    \chi_{\vec{k},E}(\vec{r} + \vec{r}_j) = e^{i\vec{p}\cdot \vec{r}}e^{z_j / \lambda} \chi_{\vec{k},E}(\vec{r}) \ .
\end{align}
Here, we have additionally taken into account the mean-free path $\lambda$ of the photoelectrons, resulting in the attenuation factor $e^{z_j / \lambda}$, which suppresses the photoemission signal from atoms far away from the surface.

Due to $\hat{\vec{p}}|\chi_{\vec{k},E}\rangle \approx \vec{p} |\chi_{\vec{k},E}\rangle $, the matrix elements in the velocity gauge simplify to 
\begin{align}
\label{eq:mel_velo_pwa}
    M_\alpha(\vec{k},E) = \vec{e} \cdot \vec{p} \sum_j e^{-i\vec{p}\cdot \vec{r}_j} e^{z_j/\lambda} C_{j\alpha}(\vec{k})
    \tilde{\phi}_j(\vec{p}) \ ,
\end{align}
where $\tilde{\phi}_j(\vec{p}) = \int d\vec{r}\, e^{-i\vec{p}\cdot\vec{r}} \phi_j(\vec{r})$ denotes the Fourier transform of the Wannier orbital $\phi_j(\vec{r})$.
The matrix elements~\eqref{eq:mel_velo_pwa} vanish for perpendicular polarization vector $\vec{e}$ and photoelectron momentum $\vec{p}$, which is a serious shortcoming of the approximation. Furthermore, there is no circular dichroism as long as $\vec{p}$ is real~\cite{moser_experimentalists_2017}. Despite the shortcomings of the PWA in the velocity gauge, the interference effects inherent to the photoemission process are already present in Eq.~\eqref{eq:mel_velo_pwa}: the phase factors $e^{-i \vec{p}\cdot\vec{r}_j}$ incorporate position of the atomic sites in the unit cell. 

Using the dipole gauge and the ACA, we can evaluate the orbital matrix elements~\eqref{eq:mel_dip_aca_orb}:
\begin{align}
    \label{eq:mel_dip_aca_orb_pw}
        M^{\mathrm{orb,ACA}}_j(\vec{k},E) = e^{-i\vec{p}\cdot \vec{r}_j} e^{z_j/\lambda} \int d\vec{r}\, e^{-i \vec{p}\cdot \vec{r}} 
    \vec{e}\cdot \vec{r}
    \phi_j(\vec{r})  \ . 
\end{align}
Inserting the orbital matrix elements~\eqref{eq:mel_dip_aca_orb_pw} into Eq.~\eqref{eq:mel_dip_aca} then yields the band-dependend photoemission matrix elements.
Similarly, the geometric phase factors $e^{-i\vec{p}\cdot \vec{r}_j}$ give rise to interference effects of different partial waves emitted from the different sub-lattice sites in the crystal. The orbital matrix elements~\eqref{eq:mel_dip_aca_orb_pw} are not necessarily zero for $\vec{e} \perp \vec{p}$, and circular dichroism is also possible (depending on the orbital character of the Wannier orbitals $\phi_j(\vec{r})$). Therefore, the combination of dipole gauge, ACA, and PWA already provide a realistic description of photoemission beyond the corresponding velocity-gauge formulation.

\subsection{Locally distorted wave approximation}

While the PWA includes basic features of the photoemission process, such as the coherent superposition of atomic transition amplitudes, many important effects are missing. In particular, the effective potential $V(\vec{r})$ has strong dips close to the atomic positions, which distorts the partial waves originating from each atom. To account for this effect, we introduce the locally distorted wave approximation (LDWA), where the photoelectron wave-function around each atomic site $\vec{r}_j$ is expanded in spherical harmonics:
\begin{align}
    \label{eq:disorted_wave_1}
    \chi_{\vec{k},E}(\vec{r}) = 4\pi \sum^\infty_{\ell = 0} \sum^{\ell}_{m=-\ell} i^\ell \frac{w_{j,\ell}(r)}{r} Y_{\ell m}(\Omega_{\vec{r}}) Y^*_{\ell m}(\Omega_{\vec{p}}) \ .
\end{align}
Here, $\Omega_{\vec{r}}$ ($\Omega_{\vec{p}}$) denotes the solid angle of the vector $\vec{r}$ ($\vec{p}$), while $Y_{\ell m}(\Omega)$ stand for the spherical harmonics. The physical picture behind the expansion~\eqref{eq:disorted_wave_1} is that the potential around each atom is spherically symmetric: $V(\vec{r}) \approx v_j(|\vec{r} - \vec{r}_j|)$ for $\vec{r} \approx \vec{r}_j$. The radial wave-functions are then determined by
\begin{align}
    \label{eq:radial_se}
    \left[-\frac12 \frac{d^2}{dr^2} + \frac{\ell(\ell + 1)}{2r^2} + v_j(r) \right] w_{j,\ell}(r) = E w_{j,\ell}(r) \ ,
\end{align}
together with the outgoing scattering boundary conditions. 

A useful special case of the LDWA is obtained by approximating the local atomic potential by an effective Coulomb potential $v_j(r) \approx - Z_j / r$. In this case, the radial wave-functions can be found explicitly~\cite{joachain_quantum_1975}:
\begin{align}
    \label{eq:coulomb_wave}
    w_{j,\ell}(r) = \frac{1}{p}\exp\left(-i\sigma_{\ell}(\eta_j)\right) F_\ell(\eta_j,p r) \ ,
\end{align}
where $\sigma_{\ell}(\eta) = \mathrm{arg}[\Gamma(\ell + 1 + i\eta)]$ is the Coulomb phase shift,  $\eta_j = -Z_j / p$ is the Sommerfeld parameter, and $F_\ell(\eta,x)$ denotes the regular Coulomb function~\cite{olver_nist_2010}. We refer to Eq.~\eqref{eq:coulomb_wave} as local Coulomb-wave approximation (LCWA). Note that the LCWA includes the PWA in the limit $\eta \rightarrow 0$, since $F_\ell(\eta, x) \rightarrow r j_{\ell}(x)$ and $\sigma_{\ell}\rightarrow 0$. In contrast to the PWA, the phase property~\eqref{eq:pw_shift} must be assumed as additional approximation to proceed. 

Assuming the validity Eq.~\eqref{eq:pw_shift}, the orbital matrix elements in the velocity gauge~\eqref{eq:mel_orbital_mom} are transformed to
\begin{align}
    \label{eq:ldwa_velo}
    M^\mathrm{orb}_j(\vec{k},E) &= 4\pi e^{-i\vec{p}\cdot\vec{r}_j} e^{z_j/\lambda_j} \sum_{\ell m} (-i)^\ell Y_{\ell m}(\Omega_{\vec{p}}) \nonumber \\ &\quad \times
    \int d\vec{r}\, \frac{w^*_{j,\ell}(r)}{r}Y^*_{\ell m}(\Omega_{\vec{r}}) \vec{e}\cdot\hat{\vec{p}} 
    \phi_j(\vec{r}) \ .
\end{align}
Similarly, we obtain the corresponding orbital matrix element within the dipole gauge and the ACA by replacing $\vec{e}\cdot \hat{\vec{p}} \rightarrow \vec{e}\cdot \hat{\vec{r}} $.

The phase of radial wave-functions $w_{j,\ell}(r)$ (which is given by $e^{i\sigma_{j,\ell}(p)}$ within the LCWA) introduces an energy-dependent phase shift between the partial waves originating from each atomic site, which is absent within the PWA. Therefore, the LDWA can reproduce full-fledged first-principle simulations or experimental spectra better than the PWA. In practice, the LCWA has been tested for a few systems in ref.~\cite{moser_toy_2023} and ref.~\cite{tanaka_spadexp_2023} with promising results.

\subsection{Radial integrals}

One of the main advantage of the Wannier representation~\eqref{eq:wannier1} is the fact that the complicated wave-like Bloch state $\psi_{\vec{k}\alpha}(\vec{r})$ can be interpreted as a superposition of quasi-atomic orbitals $\phi_j(\vec{r})$. Indeed, within empirical TB approaches such as the Slater-Koster method, the orbitals $\phi_j(\vec{r})$ are chosen as atom-like orbitals with spherical symmetry. Within the first-principle Wannierization approach, atom-like orbitals can also be imposed or obtained approximately in the framework of projective Wannier functions. Upon maximizing the projectiblity~\cite{qiao_automated_2023}, the atom-like orbitals that best span the Bloch states can be found. Either way, if we assume 
\begin{align}
    \label{eq:wannier_atomic}
    \phi_j(\vec{r}) = \frac{u_j(r)}{r} Y_{\ell_j m_j}(\Omega_{\vec{r}}) \ ,
\end{align}
the expressions for the orbital matrix elements can be simplified significantly. Furthermore, the interpretation in terms of the different cross sections of the dipole-allowed transitions becomes even more transparent.

\begin{figure*}[t]
    \centering
    \includegraphics[width=1.0\textwidth]{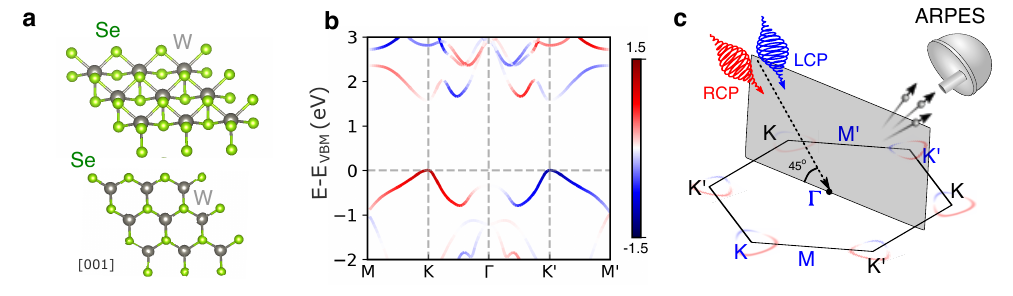}
    \caption{\textbf{a} Sketch of the crystal structure of monolayer WSe$_2$. \textbf{b}
    Band structure of monolayer WSe$_2$ (without spin-orbit interaction) in the vicinity of the valence band maximum (VBM). The color scale of the bands encodes the local orbital angular momentum along the in-coming light direction for W-5d orbitals. The k-path in follows the blue high symmetry points in \textbf{c}. \textbf{c} Geometry used in all calculations. The the angle of incidence is fixed at $\alpha = 45^\circ$ with respect to the surface along the scattering plane (shaded area).}
    \label{fig:schematic}
\end{figure*}

In the velocity gauge and using the commutation relation $\hat{\vec{p}} = -i [\hat{\vec{r}}, \hat{\vec{p}}^2]/2$, one finds
\begin{widetext}    
\begin{align}
    \label{eq:intregral_velocity}
    \int d\vec{r}\, \frac{w^*_{j,\ell}(r)}{r}Y^*_{\ell m}(\Omega_{\vec{r}}) \hat{p}_\mu
    \phi_j(\vec{r}) = -i (-1)^m C_\mu(\ell m, \ell_j m_j) \left[
    J^{(1)}_{j,\ell} + \left(\frac{\ell_j(\ell_j+1) - \ell(\ell+1)}{2}\right) J^{(2)}_{j,\ell} \ .
    \right]
\end{align}
\end{widetext}
Here we have defined the angular coefficients
\begin{align}
    \label{eq:angular coefficients}
    C_\mu(\ell_1 m_1, \ell_2 m_2) = \sqrt{\frac{4\pi}{3}} \sum^{1}_{s-1} u^s_\mu \begin{bmatrix} \ell_1 & 1 & \ell_2 \\
    -m_1 & s & m_2\end{bmatrix}
\end{align}
with the vectors $\vec{u}^{-1} = (1,i,0)^T/\sqrt{2}$, $\vec{u}^{0} = (0,0,1)^T$, $\vec{u}^{+1} = (1, -i, 0)^T /\sqrt{2}$, while the matrix in square brackets denote the Gaunt coefficients~\cite{joachain_quantum_1975}. The only information about the radial wave-function $R_j(r)$ is encoded in the radial integrals
\begin{align}
    \label{eq:radint_velo}
    J^{(1)}_{j,\ell} = \int^\infty_0 dr\, w^*_{j,\ell}(r) u'_j(r) \ , \
    J^{(2)}_{j,\ell} = \int^\infty_0 dr\, w^*_{j,\ell}(r) \frac{1}{r} u_j(r) \ .
\end{align}

The derivation for the dipole gauge is similar. One finds
\begin{align}
    \label{eq:intregral_dipole}
    \int d\vec{r}\, \frac{w^*_{j,\ell}(r)}{r}Y^*_{\ell m}(\Omega_{\vec{r}}) \hat{e}_\mu
    \phi_j(\vec{r}) = (-1)^m C_\mu(\ell m, \ell_j m_j) I_{j,\ell}
\end{align}
with the radial integral
\begin{align}
    \label{eq:radint_dipole}
    I_{j,\ell}= \int^\infty_0 dr\, w^*_{j,\ell}(r) r u_j(r) \ .
\end{align}
Substituting either Eq.~\eqref{eq:intregral_dipole} or \eqref{eq:intregral_velocity}
into Eq.~\eqref{eq:ldwa_velo} then yields the orbital matrix elements.

As in atomic physics, the dipole selection rule (encoded in the coefficients~\eqref{eq:angular coefficients}) constrains the angular momentum of the outgoing wave to $\ell = \ell_j \pm 1$, while the magnetic quantum number $m = m_j, m_j \pm 1$ depending on the polarization. Thus, the orbital matrix element~\eqref{eq:ldwa_velo} is fully characterized by the integrals $J^{(1)}_{j,\ell_j \pm 1}$ and $J^{(2)}_{j,\ell_j \pm 1}$ in velocity gauge, and by $I_{j,\ell_j \pm 1}$ in dipole gauge. 

In practice, the radial integrals~\eqref{eq:radint_velo} or \eqref{eq:radint_dipole} can be computed from the Wannier orbitals (or the closest approximation by atom-like orbitals), providing a first-principle framework to computing the photoemission matrix elements within the various approximations to the final states. Alternatively, one can treat the integrals as fitting parameters to model photoemission spectra obtained from experiments.

\subsection{Gauge invariance considerations}

While the equivalence of dipole gauge in velocity gauge is broken by invoking the ACA, the orbital matrix elements are still equivalent if the same partial-wave expansion for the final states $w_\ell(r)$ is used around each atom. This equivalence is an expression of \emph{local} gauge invariance. Consider the purely atomic problem with a spherically symmetric potential $v(r)$. If both the bound states $u_j(r)$ and the final states $w_\ell(r)$ obey the same radial Schr\"odinger equation,
\begin{align}
    \label{eq:radialse}
    & \left[-\frac12 \frac{d^2}{dr^2} + \frac{\ell_j(\ell_j+1)}{2r^2} + v(r)\right] u_j(r) = \varepsilon_j u_j(r) \nonumber \\ 
    & \left[-\frac12 \frac{d^2}{dr^2} + \frac{\ell(\ell+1)}{2r^2} + v(r)\right] w_\ell(r) = \varepsilon w_\ell(r) \ ,
\end{align}
the radial integrals in both gauges are related by
\begin{align}
    \label{eq:gauge_con}
    J^{(1)}_{j,\ell} + \left(\frac{\ell_j(\ell_j+1) - \ell(\ell+1)}{2}\right) J^{(2)}_{j,\ell}
    = (\varepsilon_j - \varepsilon) I_{j,\ell} \ .
\end{align}
The constraint~\eqref{eq:gauge_con} is useful for ensuring consistency between the gauge choices, which is especially important when invoking additional approximations to the partial waves of the photoelectron states $w_\ell(r)$. As an important application, Eq.~\eqref{eq:gauge_con} allows us to choose the effective charges $Z_j$ within the LCWA in a consistent manner. Assuming we have access to the orbital energies $\varepsilon_j$ and (a guess of) the radial wave-functions $u_j(r)$, $Z_j$ can be chosen such that the violation of Eq.~\eqref{eq:gauge_con} is minimal. We demonstrate the power of this approach below.

\begin{figure*}[t]
    \centering
    \includegraphics[width=0.9\textwidth]{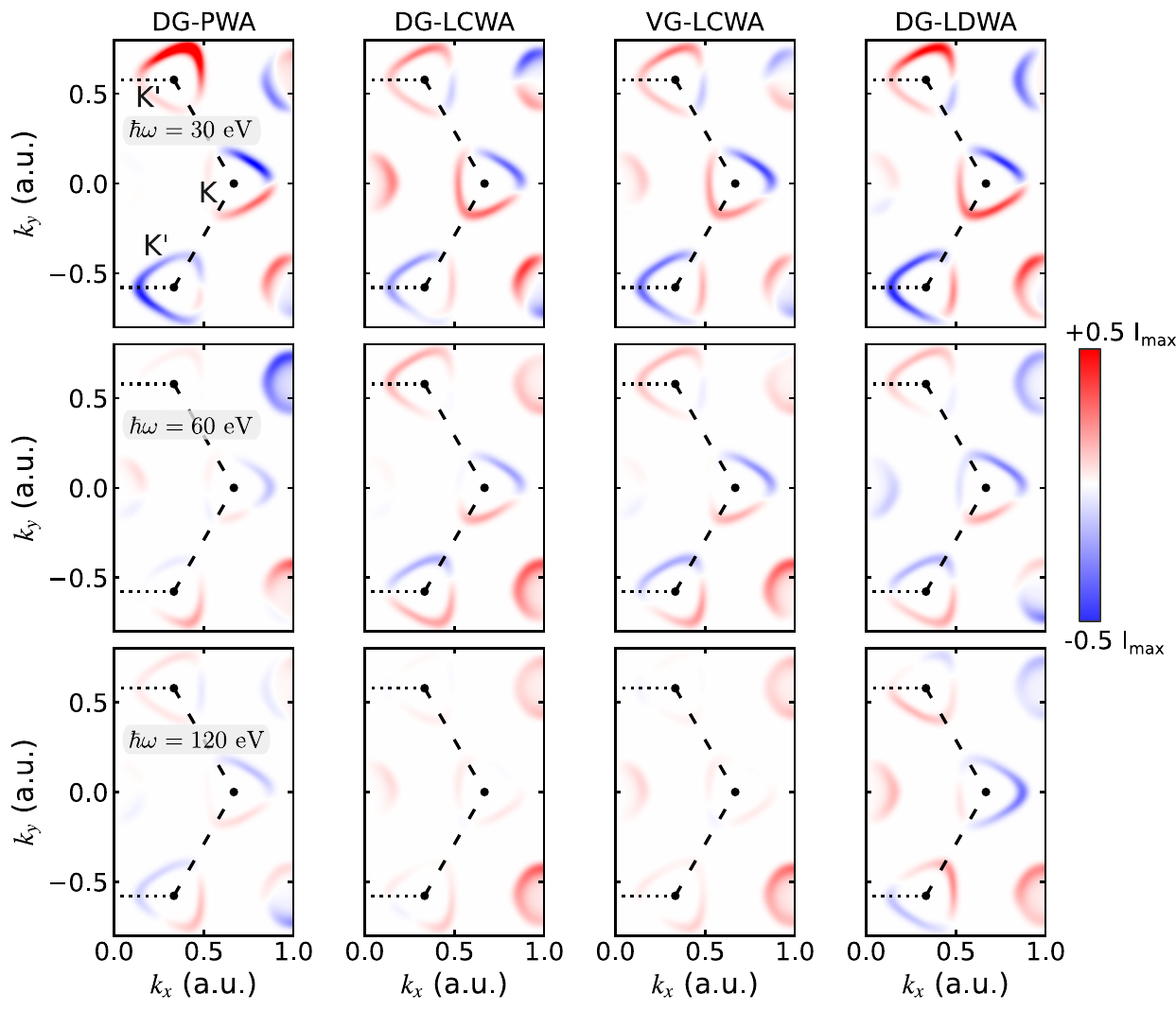}
    \caption{Simulated circular dichroism in the angular distribution (CDAD) $I_\mathrm{CDAD}(\vec{k},E)$ for monolayer WSe$_2$ at binding energy $E - E_\mathrm{VBM} = -0.4$~eV with respect to the valence band maximum (VBM). The different panels show the results at different photon energies (indicated in the panels on the left-hand side), obtained within the approximations discussed in the main text: dipole gauge and plane-wave approximation (DG-PWA), dipole gauge and local Coulomb-wave approximation (DG-LCWA), velocity and local Coulomb-wave approximation (VG-LCWA), and dipole gauge and distorted-wave approximation (DG-LDWA). The signal is anti-symmetric upon reversing the sign of $k_x$. The CDAD has been normalized to the maximum unpolarized intensity $I_\mathrm{max}$.}
    \label{fig:cuts}
\end{figure*}

\section{Illustrative example: monolayer tungsten diselenide}

To show how the various light-matter coupling schemes and approximations work in practice, we have performed calculations for monolayer tungsten diselenide WSe$_2$. This compound belongs to the class of two-dimensional (2D) materials, which are currently in the spotlight for the interplay of lattice, spin and orbital degrees of freedom, strongly bound excitons, and strong light-matter coupling.  The crystal structure of monolayer WSe$_2$ is shown in Fig.~\ref{fig:schematic}\textbf{a}, with one tungsten and two selenium atoms per unit cell. 

We focus on the approximations to the photoemission matrix elements and treat the equilibrium electronic structure from first principles. To this end, we performed density-functional theory (DFT) calculations using the \textsc{Quantum Espresso} software package~\cite{giannozzi_quantum_2009}. We used the PBE functional and pseudopotentials from the \textsc{PseudoDojo} project~\cite{van_setten_pseudodojo_2018}. The pseudopotential also provides us with the pseudo-atomic radial wave-functions $u_j(r)$, and an optimal basis $\phi_j(\vec{r})$ to represent the Bloch states $\psi_{\vec{k}\alpha}(\vec{r})$. We neglect spin-orbit interaction for simplicity, since it does not influence the OAM texture near the top of the valence band.
The Wannier representation \eqref{eq:wannier1} is obtained by a modified version of the \textsc{Wannier90} package~\cite{pizzi_wannier90_2020}, which allows us to construct the Wannier Hamiltonian $H^{(\mathrm{W})}_{j j^\prime}(\vec{k})$ in the basis of the pseudo-atomic basis. With this Wannier Hamiltonian, we calculate the band structure and the OAM of the W-5d orbitals, which are presented in Fig.~\ref{fig:schematic}\textbf{b}. Due to time-reversal symmetry, the OAM is antisymmetric with respect to the $\Gamma$ point along with indicated path in the BZ.

With this first-principle input, we have computed the photoemission matrix elements using the various methods presented in Section~\ref{sec:wannier_method}. To employ the LDWA, we have solve the radial Schr\"odinger equation~\eqref{eq:radialse} using the all-electron potential of the isolated W and Se atoms, respectively, providing us with the atomic photoelectron states $w_\ell(r)$. To test the LCWA, we have replaced $w_\ell(r)$ by Eq.~\eqref{eq:coulomb_wave} and evaluated the violation of the gauge constraint~\eqref{eq:gauge_con} as a function of $Z_j$. Mimizing the violation of Eq.~\eqref{eq:gauge_con} yields $Z_d \approx 3$ for the W-$5d$ orbitals and $Z_p \approx 2.5$ for the Se-$4p$ orbitals, which dominate the orbital character of the top valence bands. 

\subsection{Circular dichroism in the angular distribution}

We calculated the photoemission signal through Fermi's Golden rule~\eqref{eq:fermi_golden_2} for a typical experimental geometry sketched in Fig.~\ref{fig:schematic}\textbf{c} for various photon energies ranging from $10$~eV to $120$~eV. Similar to previous experimental studies~\cite{beaulieu_revealing_2020-1,cho_studying_2021} we assume the scattering plane to align with mirror plane of the crystal.
As characteristic observable we focus o the circular dichroism in the angular distribution (CDAD)
\begin{align}
    \label{eq:cdad}
    I_\mathrm{CDAD}(\vec{k},E) = I_\mathrm{LCP}(\vec{k},E) - I_\mathrm{RCP}(\vec{k},E)  \ ,
\end{align}
where $I_\mathrm{LCP}(\vec{k},E)$ ($I_\mathrm{RCP}(\vec{k},E)$) denotes the intensity using left-hand (right-hand) circularly polarization photons. Due to the mirror symmetry, which is preserved by the geometry of the simulations, the CDAD is anti-symmetric with respect to the mirror plane: $I_\mathrm{CDAD}(k_x, k_y, E) = -I_\mathrm{CDAD}(-k_x, k_y, E)$.

To find potential signatures of the pronounced OAM of the Bloch electrons of the top valence band, we focus on the binding energies close the valence band maximum (VBM). In Fig.~\ref{fig:cuts} we present the simulated CDAD for different photon energies. We also show results obtained using the different gauges and approximations discussed in Section~\ref{sec:wannier_method}. 

Inspecting the different approximations for the photon energy $\hbar\omega=30$~eV, we find overall qualitative agreement among the different methods. All of them predict a polar texture of the CDAD around the K point, with a nodal line separating positive and negative CDAD at the same position. The DG-PWA predicts a weaker CDAD than the other approximations, and fails to reproduce the sign changes at the K' points. Increasing the photon energy, the nocal line shifts and the CDAD at K becomes negative. While this trend is captured by all approximations, the DG-PWA shows large deviations from the DG-LDWA (that includes the scattering phases). Except for the largest photon energy $\hbar\omega=120$~eV, the DG-LCWA and the VG-LCWA (which are consistent among each other) are in good agreement with the DG-LDWA. This behavior illustrates that enforcing minimal violation of the local gauge constraint~\eqref{eq:gauge_con} yields a very good approximation if the precise form of the atomic potential is unknown.   

\begin{figure}[t]
    \centering
    \includegraphics[width=\columnwidth]{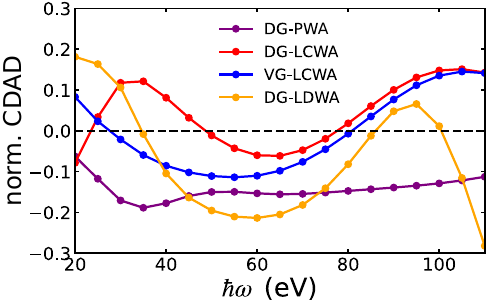}
    \caption{Photon-energy dependence of the valley-integrated CDAD $N_\mathrm{CDAD}$, at fixed binding energy $E - E_\mathrm{VBM} = -0.4$~eV, in the different approximations.}
    \label{fig:photonenergy}
\end{figure}

\subsection{Photon energy dependence of the circular dichroism}

Now we inspect the photon-energy dependence more closely. We computed the valley-integrated intensity $I^{\mathrm{av}}_{K} = \int d\vec{k} (I_\mathrm{LCP}(\vec{k},E) + I_\mathrm{RCP}(\vec{k},E))$ (with a integration box large enough to include the intensity around K as shown in Fig.~\ref{fig:cuts}) and the CDAD $I^{\mathrm{CDAD}}_{K} = \int d\vec{k} (I_\mathrm{LCP}(\vec{k},E) - I_\mathrm{RCP}(\vec{k},E))$. In Fig.~\ref{fig:photonenergy} we show the normalized valley-integrated CDAD $N_\mathrm{CDAD} = I^{\mathrm{CDAD}}_{K} / I^{\mathrm{av}}_{K}$ as a function of the photon energy. While all approximations predict oscillatory features, the behavior DG-PWA deviates from the other approximations, as no sign change of $N_\mathrm{CDAD}$ is predicted. The DG-LCWA and the VG-LCWA are consistent with each other at larger photon energy, but show some deviations in range $\hbar\omega=20$~eV to $\hbar\omega = 60$~eV. Both of these approximations capture the sign change of the CDAD at roughly the same photon energy as predicted by the DG-LDWA. The main difference between the DG-LDWA and the other approximations is the 
missing scattering phase, since all radial integrals $I_{j,\ell}$ are purely real. Furthermore, the photon-energy dependence is sensitive to the relative strength of the radial integrals, which determine the relative contribution of the various channels of the final states.

\begin{figure}[t]
    \centering
    \includegraphics[width=\columnwidth]{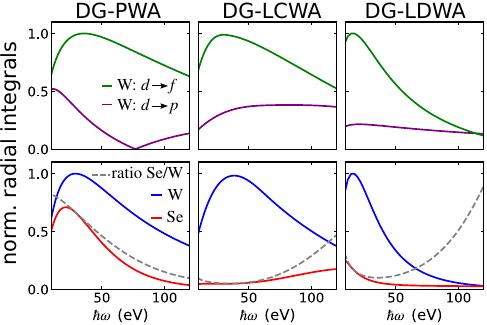}
    \caption{Absolute value of the radial integrals $|I_{j,\ell}|$ (normalized by their respective maximum). Top panels: integrals for the W atom, comparing the $d\rightarrow f$ and the $d\rightarrow p$ final channels. Bottom panels: atom-resolved total radial integrals $I^\mathrm{Se}$, $I^\mathrm{W}$ for the W and the Se atoms. The gray dashed line represents the ratio $I^\mathrm{Se}/I^\mathrm{W}$.}
    \label{fig:crosssections}
\end{figure}

In Fig.~\ref{fig:crosssections} we inspect the normalized absolute value $|I_{j,\ell}|$ for the dipole gauge. We start with the W atoms, since the W-$d_{\pm 2}$ orbitals have the largest weight at the VBM.
Within all approximations to the final states, the dipole-allowed $d\rightarrow f$ channel ($\ell = 3$ in the corresponding integrals) is the most important for the W atoms. The ratio is however different at low photon energies within the PWA, in contrast to the distorted-wave models. Taking the realistic local potential into account, the DG-LDWA predicts a cross-over between $d\rightarrow f$ channel and the $d\rightarrow p$ at $\hbar\omega \approx 120$~eV, which coincides with the rapid sign change of the CDAD in Fig.~\ref{fig:photonenergy}. The DG-LCWA shows a similar trend, albeit the crossing point occurs outside the considered photon energy range. 

It is also interesting to investigate the behavior of the Se atoms, since the Bloch states gain some appreciable weight of Se-$4p$ orbitals at binding energy $E-E_\mathrm{VBM}=-0.4$~eV studied in Fig.~\ref{fig:cuts}. We can re-label the orbital indices $j=(a,m)$, where $m$ is the magnetic quantum number corresponding to the angular momentum of each orbital, while $a\in \{\mathrm{W}, \mathrm{Se}_1, \mathrm{Se}_2 \}$ denotes the atoms in the unit cell.
Using this notation, we define the atom-resolved total contributions as $I^\mathrm{Se} = \sum^{1}_{m=-1} \sum_{\ell=0,2} |I_{m,\ell}|$ and $I^\mathrm{W} = \sum^2_{m=-2}\sum_{\ell=1,3} |I_{j,\ell}|$, where we focus on the Se-$4p$ and the W-$5d$ orbitals.
While the total radial integral $I^\mathrm{W}$ is always bigger than $I^\mathrm{Se}$ within all approximations, the ratio $I^\mathrm{Se}/I^\mathrm{W}$ is differs considerably. The DG-PWA predicts a drop of the ratio, starting from a high value of $I^\mathrm{Se}/I^\mathrm{W}\approx 0.8$, while the DG-LCWA and the DG-LDWA show a rise towards higher photon energies. The consistency of DG-LCWA and DG-LDWA is another indication that the gauge constraint~\eqref{eq:gauge_con} provides a realistic description for the radial integrals.

\subsection{Interference effects}

From the discussion of the final state channels it becomes clear that both the W and the Se atoms contribute to the photoemission signal and the CDAD in particular. This is especially important to explain the oscillatory photon energy dependence observed in Fig.~\ref{fig:photonenergy}, which is clear indication for interference effects.

With the Wannier-ARPES models we can analyze the various interference channels in detail. To this end we rewrite the CDAD directly in terms of the orbital matrix elements $M^\mathrm{orb}_j(\vec{k},E)$ and the Wannier coefficients $C_{j\alpha}(\vec{k})$. After collecting all terms, we find the compact form
\begin{align}
    \label{eq:tensorrep}
    I_{\mathrm{CDAD}}(\mathbf{k},E) & = \sum_{j,j^\prime} C^*_{j\alpha}(\mathbf{k}) T_{j j^\prime}(\mathbf{k},E) C_{j^\prime\alpha}(\mathbf{k}) \ .
\end{align}
Here, the $T_{j j^\prime}(\mathbf{k},E)$ includes the matrix elements $M^\mathrm{orb}_{j,(\mathrm{LCP})}(\mathbf{k},E)$ ($M^\mathrm{orb}_{j,(\mathrm{RCP})}(\mathbf{k},E)$ ) with respect to LCP (RCP) photons. The advantage of the representation~\eqref{eq:tensorrep} is that we can now separate out the initial states $j=(a,m)$. The most important terms are associated to the W-$5d$ and the Se-$4p$ orbitals, such that we can approximate

\begin{widetext}
\begin{align}
    \label{eq:cd_simplified}
    I_{\mathrm{CDAD}}(\mathbf{k},E) &\approx \sum^{2}_{m,m^\prime=-2} C^*_{(\mathrm{W},m)\alpha}(\mathbf{k}) T^{(5d)}_{m m^\prime}(\mathbf{k},E) C_{(\mathrm{W},m^\prime)\alpha}(\mathbf{k}) +\Big(\sum^2_{m=-2}\sum^{1}_{m^\prime=-1} C^*_{(\mathrm{W},m)\alpha}(\mathbf{k}) T^{(5d)-(4p),1}_{m m^\prime}(\mathbf{k},E) C_{(\mathrm{Se}1,m^\prime)\alpha}(\mathbf{k}) + h.c.\Big) \nonumber\\
    &+ \Big(\sum^2_{m=-2}\sum^{1}_{m^\prime=-1} C^*_{(W,m)\alpha}(\mathbf{k}) T^{(5d)-(4p),2}_{m m^\prime}(\mathbf{k},E) C_{(\mathrm{Se}2,m^\prime)\alpha}(\mathbf{k}) + h.c. \Big)\ ,
\end{align}
\end{widetext}
where $h.c.$ stands for the hermitian conjugate. The first term in Eq.~\eqref{eq:cd_simplified} describes the contributions from the W atoms only, while the second (third) term incorporates the interference between the partial waves originating from the W and the first (second) Se atom in the unit cell. The Se-Se contribution has been neglected here. For simplicity, we computed the tensors within the DG-PWA, which captures the essential features of the interference effects discussed here.

Now we can address the question of how the local OAM shown in Fig.~\ref{fig:schematic}\textbf{b} is reflected in the CDAD. First we inspect the local OAM of each atom $a$
\begin{align}
    \label{eq:local_OAM_def}
    \vec{L}(\vec{k}) = \sum_{m m^\prime} C^*_{a, m}(\vec{k}) \vec{L}_{m m^\prime} C_{a, m^\prime}(\vec{k}) \ ,
\end{align}
where $\vec{L}_{m m^\prime}$ denotes the matrix representation of the angular momentum operator for angular momentum $\ell=2$ for $a = \mathrm{W}$ and $\ell=1$ for $a = \mathrm{Se}$. For a direct comparison with the CDAD it is convenient to inspect the OAM density,
\begin{align}
 \label{eq:local_OAM_dens}
    \vec{L}(\vec{k},E) = \vec{L}(\vec{k}) g(\varepsilon_\alpha(\vec{k}) + \omega - E) \ ,
\end{align}
where $g(\varepsilon)$ denotes a Gaussian function with the width as used for the photoemission calculations.

\begin{figure}[t]
    \centering
    \includegraphics[width=\columnwidth]{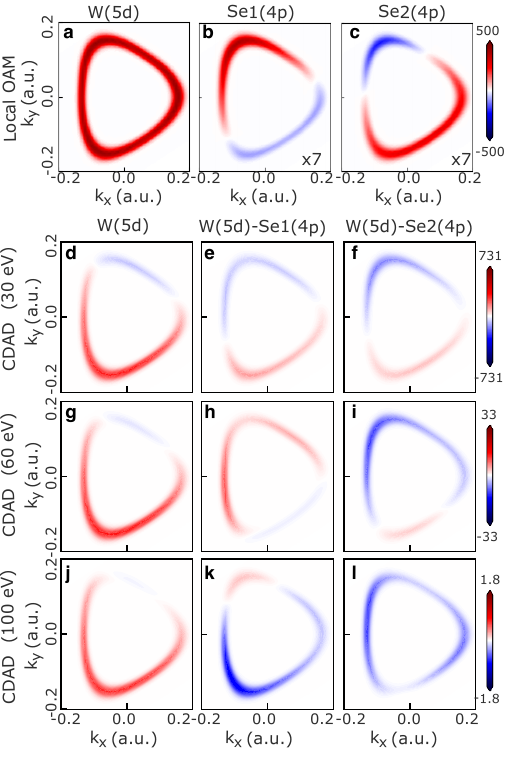}
    \caption{\textbf{a}-\textbf{c} Local OAM density for W-5d, Se1-4p, and Se2-4p orbitals, respectively. The local OAM density for Se-4p orbitals is re-scaled for better visibility. \textbf{d}-\textbf{f} Simulated CDAD at photon energy 30 eV from local contribution of W, interference between W-Se1, and interference between W-Se2. \textbf{d}-\textbf{f} Simulated CDAD at photon energy 60 eV. \textbf{j}-\textbf{l} Simulated CDAD at photon energy 100 eV. }
    \label{fig:loam_cd}
\end{figure}

In Fig.~\ref{fig:loam_cd}\textbf{a}--\textbf{c} we show the local OAM density~\eqref{eq:local_OAM_dens} projected onto the incidence direction of the photons for the three atoms in the unit cell. The OAM of the W-$5d$ atoms clearly dominates, since the Bloch states at the VBM are mostly composed of the magnetic $d_{\pm 2}$ (at K/K') orbitals. Now we inspect the different contributions to the CDAD in Eq.~\eqref{eq:cd_simplified}. The contribution originating from the W atoms only reflects the local OAM, as a clear positive CDAD is observed. However, even if interference effects with the Se atoms are switched off, the CDAD is not entirely positive. This behavior is due to the properties of the tensor $T_{mm^\prime}^{(5d)}(\vec{k},E)$. Switching the quantization axis to be parallel to the direction of incidence, a one-to-one correspondence of OAM and this local CDAD contribution would only be expected if all off-diagonal elements $T_{mm^\prime}^{(5d)}(\vec{k},E)$ for $m\ne m^\prime$ were zero, or, in other words, if the \emph{intra-atomic} interferences were absent. This is however not the case at lower photon energies (which is an intrinsic property of $d$ orbitals~\cite{yen_controllable_2023}, in contrast to $p$ orbitals~\cite{moser_toy_2023}). Only at larger photon energies the off-diagonal elements vanish, and a positive contribution to the CDAD can be expected (Fig.~\ref{fig:loam_cd}\textbf{d},\textbf{g},\textbf{j}). 

Even the contribution of the W-$5d$ orbitals to the CDAD does not provide a one-to-one correspondence to the OAM in the XUV regime studied here. Including the remaining interference terms in Eq.~\eqref{eq:cd_simplified} further obscures this link. In Fig.~\ref{fig:loam_cd}\textbf{e}, \textbf{h}, \textbf{k} (\textbf{f}, \textbf{i}, \textbf{l}) we show the interference terms originating from the first (second) Se atom. The interference contributions exhibit a strong dependence on the photon energy, which is of mostly geometric origin. The orbital matrix elements $M^\mathrm{orb}_j(\vec{k},E)$ contain the structure factor $e^{-i\vec{p}\cdot \vec{r}_j} = e^{-i\vec{p}\cdot \vec{r}_a}$. As a consequence, the tensors in~\eqref{eq:cd_simplified} are proportional to the structure factor $T^{(5d)-(4p)1,2}_{m m^\prime}(\vec{k},E) \propto \exp[-i\vec{p}\cdot \vec{v}^{(1,2)}_{a}]$, where $\vec{v}^{(1,2)}_{a}$ is the vector pointing from the W atom to Se atom 1,2. While in-plane component is fixed $\vec{p}_\parallel=\vec{k}$, the out-of-plane component $p_\perp$ depends on the photon energy: $p_\perp = \sqrt{2 E - \vec{k}^2} = \sqrt{2(\varepsilon_\alpha(\vec{k}) + \omega) - \vec{k}^2}$. As a result, the tensors $T^{(5d)-(4p)1,2}_{m m^\prime}(\vec{k},E)$ are proportional to a photon-energy dependent phase factor that impacts the CDAD directly. Summing the the three contributions in Fig.~\ref{fig:loam_cd}\textbf{d}--\textbf{f} reproduces the CDAD in the vicinity of K in Fig.~\ref{fig:cuts} for $\hbar\omega=30$~eV. Similarly, summing the terms corresponding to Fig.~\ref{fig:loam_cd}\textbf{j}--\textbf{l} explains the suppression of the CDAD at larger photon energy observed within the DG-PWA.

\section{Conclusions}

We have presented a detailed derivation and discussion of the various model approaches to calculating ARPES intensities. Accurately representing the Bloch wave-functions in terms of quasi-atomic Wannier functions allowed us to highlight the role of approximating the final states in combination with choosing the light-matter gauge. In case the potential $V(\vec{r})$ around each atom is available, one can directly compute the locally distorted waves, yielding realistic radial integrals and phase shifts.
Comparison to experiments~\cite{yen_controllable_2023} shows that complementing the resulting LDWA with such first-principle input provides a very accurate description of ARPES, as long as multi-band effects in the final states do not play an important role. As good proxy for the local potential we suggested to compute the distorted waves for Kohn-Sham potential of the isolated atoms, which will not deviate much from the potential in the crystal. The quality of this approximation can be assessed by inspecting how well the Bloch wave-function can be represented in the basis of isolated atoms. For our benchmark system, monolayer WSe$_2$, this criterion is fulfilled with high accuracy. 

Taking the LDWA as a benchmark, we tested the different other approximations. While the popular PWA provides a reasonable overall description, it fails to reproduce the correct photon-energy dependence. This shortcoming is explained by the incorrect behavior of the radial integrals, predicting the wrong energy dependence of the ratio of contributions from W and Se atoms. Interestingly, the approximating the final states as local Coulomb waves around each atom yields results comparable to the first-principle LDWA, provided the violation of the local gauge invariance condition is minimized. This principle is expected to be applicable generally, and should be the preferred approximation in case the local potential around each atom is not available. 

We then utilized the main advantage of Wannier-ARPES models as compared to one-step calculations: the ability to distenangle and understand the various contributions to the total signal. In particular, we highlighted the ubiquitous interference effects in the CDAD. Besides intra-atomic inteference, which already complicates the signal if $d$ orbitals are involved, inter-atomic interference plays the main role for the photon-energy dependence of the CDAD observed discussed here (and observed in experiments~\cite{scholz_reversal_2013,fedchenko_4d_2019,yen_controllable_2023}).

With the inevitable interferences, how can one link the OAM as property of the Bloch wave-function to the CDAD? In general, a one-to-one correspondence can not be established. Nevertheless, the interference analysis presented in this work enables a few possible strategies. First of all, inter-atomic interference only plays a role if (1) the relevant Bloch states are comprised of orbitals localized at different lattice sites, and (2) the cross sections of these orbitals are comparable. Since at high photon energies one atomic cross section typically dominates, performing experiments in the X-ray range reduces the inteferences. This strategy works for systems like 
WSe$_2$, where the relative contribution of the single W atom per unit cell can be enhanced. For materials with multiple atoms of the same species in the unit cell, the inter-atomic interference can not be eliminated even at large photon energy. A more promising, albeit material-specific, strategy is to exploit crystal symmetries. For instance, rotating WSe$_2$ by 60$^\circ$ emulates a time-reversal operation~\cite{beaulieu_revealing_2020-1}. Taking the difference of the circular (or linear) dichroism partially removes interference terms~\cite{schusser_towards_2024}, as the anti-symmetric contribution is enhanced. A third option is to identify specific features in the CDAD that remain robust when changing the photon energy, as demonstrated in ref.~\cite{yen_controllable_2023}.

Ultimately, combining experiments with accurate modeling, as presented in this paper, is most promising route to extracting detailed information from ARPES measurements.Disentangling the various intereference channels greatly facilitates the separation of intrinsic material properties and extrinsic effects. Combining the Wannier-ARPES methods presented in this work with first-principle input provides an excellent tool for zeroing in on this goal.

\section*{Acknowledgments}

This research was supported by the NCCR MARVEL, a National Centre of Competence in Research, funded by the Swiss National Science Foundation (grant number 205602).
M.S. and Y.Y. acknowledge support from SNSF Ambizione Grant No. PZ00P2 193527. 

%

\end{document}